# Tailoring ferromagnetism through electrically controlled morphology

*Giovanni Vinai,*[\*,†] *Federico Motti,*[†] *Valentina Bonanni,*[†,||] *Damiano Cassese,*[‡] *Stefano Prato,*[‡] *Giorgio Rossi,*[†,||] *Giancarlo Panaccione,*[†] *Piero Torelli*[\*,†]

[†]Laboratorio TASC, IOM-CNR, S.S. 14 km 163.5, Basovizza, I-34149 Trieste, Italy

[||]Department of Physics, Università degli Studi di Milano, Via Celoria 16, I-20133 Milano, Italy

[‡] A.P.E. Research Srl, AREA Science Park, Basovizza, Trieste, Italy

*Correspondence to: vinai@iom.cnr.it, piero.torelli@elettra.eu

**ABSTRACT**

Converse magnetoelectric coupling in artificial multiferroics is generally modelled through three possible mechanisms: charge transfer, strain mediated or ion migration. Here we demonstrate a novel and highly reliable approach, where electrically controlled morphological modifications control the ferromagnetic response of a magnetoelectric heterostructure, specifically $Fe_xMn_{1-x}$ ferromagnetic films on ferroelectric PMN-PT (001) substrates. The ferroelectric PMN-PT substrates present, in correspondence to electrical switching, fully reversible morphological changes at the surface, to which correspond reproducible modifications of the ferromagnetic response of the $Fe_xMn_{1-x}$ films. Topographic analysis by atomic force microscopy shows the formation of surface cracks after application of a positive electric field up to 6 kV/cm, which disappear after application of negative voltage of the same amplitude. In-operando x-ray magnetic circular dichroic spectroscopy at Fe edge in $Fe_xMn_{1-x}$ layers shows local variations of dichroic signal up to a factor 2.5 as a function of the electrically-driven morphological state. These findings highlight the role of morphology and surface topography as a key aspect in magnetoelectric coupling, whose proof of electrically reversible modification of the magnetic response adds a new possibility in the design of multiferroic heterostructures with electrically controlled functionalities.

**MAIN TEXT**

In recent years, the possibility of controlling the magnetization of a thin film with an electric field has driven a tremendous research effort.[1–8] Unfortunately, the materials that display a natural magnetoelectric coupling, known as multiferroics,[9–11] are scarce in nature and generally display weak ferromagnetic response.[7,8,12] The absence of bulk multiferroics with the desired properties has pushed towards the investigation of heterostructures in which two components with different ferroic orders are coupled through an interface to obtain enhanced performances.[11] The most typical example of this process is represented by ferromagnetic films deposited onto ferroelectric substrates to obtain high magnetoelectric coupling coefficients.[13–31] Until now this path has proven to be promising, leading to the discovery of systems with very high magnetoelectric coupling coefficient and of new effects such as electrically driven magnetic phase transitions. Three main mechanisms are reported in literature as responsible of the interfacial magnetoelectric coupling: charge accumulation/depletion at the interface, strain-mediated effects and ion migration.[1–5] Such magnetoelectric interactions can be maximized by exploiting specific characteristics of the materials, such as strongly correlated magnetic oxides,[13,15,28,32,33] mixed-phase metallic alloys,[16–

[19] structural transitions[20–25] or nanostructuration.[26,27,34,35] In this contest, topographical studies of the heterostructure are relatively rare,[27,28,35–38] despite the great variety of materials used in both ferroelectric and ferromagnetic layers. More importantly, they do not take into account possible modifications of the morphology after ferroelectric transitions.

Here we present evidence of a new mechanism responsible of converse magnetoelectric coupling. By combining x-ray magnetic dichroism (XMCD) and atomic force microscopy, we observed a correlation between electrically induced mesoscopic morphological transitions and magnetic response. Specifically, samples showed the appearance and annihilation of surface cracks by electrically switching the ferroelectric state of the substrate; the two morphological states display significantly different magnetic properties, which are fully reversible after ferroelectric switch.

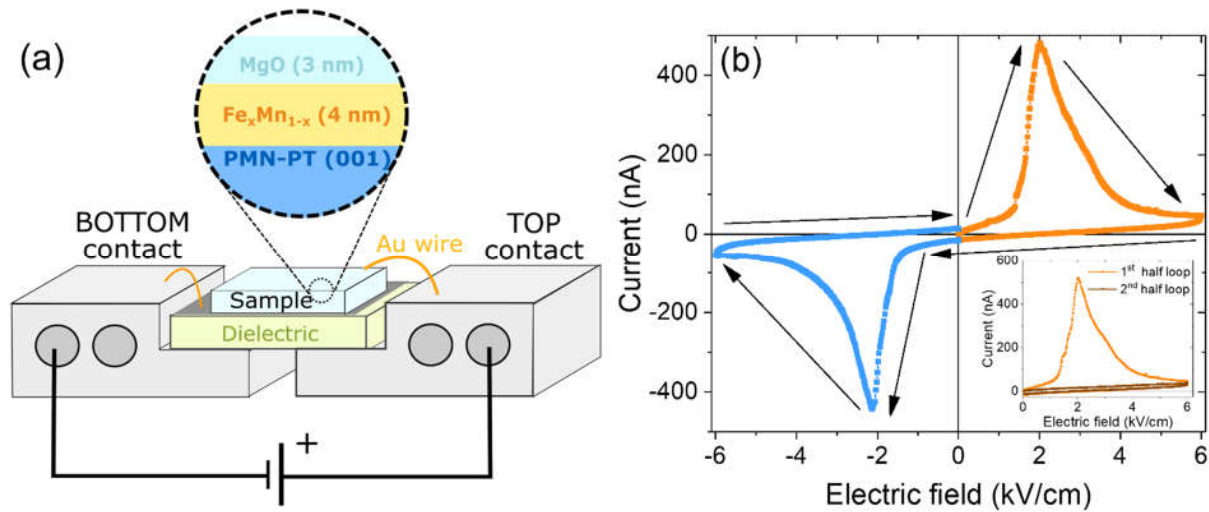

**Figure 1.** (a) Experimental setup for sample electrical switching and characterization. In the inset, schematic of the sample stack. (b) I(E) curve with PMN-PT electrical switching under both positive and negative applied bias. In the inset, I(E) curves before and after positive electrical switching.

To this scope a series of 4 nm thick $Fe_xMn_{1-x}$ films were deposited on $0.6Pb(Mg_{1/3}Nb_{2/3})O_3$–$0.4PbTiO_3$ (PMN-PT) (001) single crystal substrates by molecular beam epitaxy (MBE) via co-deposition from Fe and Mn evaporators. FeMn thin films were then capped by 3 nm MgO protecting layer (Figure 1); x ratio ranged from 0.5 to 1 in order explore the magnetic phase diagram in both the antiferromagnetic (AFM) and ferromagnetic (FM) phases.[39–41] This information is important because in the proximity of the critical point (i.e. the AFM to FM transition) the weak ferromagnetic interactions lead to a maximization of the magnetoelectric coupling.[17–19] A 4 nm thickness was chosen in order to be sensitive to the interfacial coupling with PMN-PT substrate,[29] while maintaining a robust ferromagnetic response at room temperature.[42]

In order to be able to measure the magnetic, chemical and topographic characteristics of FeMn thin films as a function of PMN-PT electrical switch state, we used a specific set of sample holders, as shown in Figure 1a. The sample is placed on a dielectric substrate, with conductive silver paint contacting the bottom part of the sample to one of the two extremities of the sample holder (bottom contact). The sample surface is then electrically contacted to the opposite contact (top contact) through a gold wire. In this way a capacitor structure is realized in which the FeMn film acts as top contact and the silver paint layer as bottom contact. The two contacts are then connected to a picoammeter/voltage source, which allows applying an out-of-plane electric voltage through the thickness of the sample up to ± 300 V (6 kV/cm), thus reversing the electrical polarization of the substrate without removing the sample from the measurement position. Figure 1b shows an example of an I(E) curve of $MgO/Fe_{85}Mn_{15}/PMN$-PT sample. The electric field is applied through the thickness of the sample, i.e. through the 0.5 mm thickness of the PMN-PT(001) substrate, linearly increasing the value from 0 to 6 kV/cm, then to -6 kV/cm, and finally back to 0. When the positive bias ($P_{up}$) is applied to the top contact (orange curve in Figure 1b), a peak of current passing through the sample is measured in correspondence of the electrical switching of the substrate; the same is observed in case of negative polarization (P$_{down}$, blue curve). The transition takes place at similar absolute values, i.e. ~ ± 2 kV/cm, for both polarities on all samples with fully reproducible and stable curves, with no sign of sample deterioration after several switches, consistently with what reported for similar substrates.[30,43] If after one electrical switching the same half electric loop is repeated afterwards (inset of Figure 1b), no current peak is detected, sign of a full electric polarization retention. The electrical switch proved to be extremely stable in time: repeating the same half electric loop several hours after the switch gave no sign of electrical peak. This demonstrates that the time stability of the PMN-PT substrate largely exceeds the XMCD and atomic force microscopy measuring time.

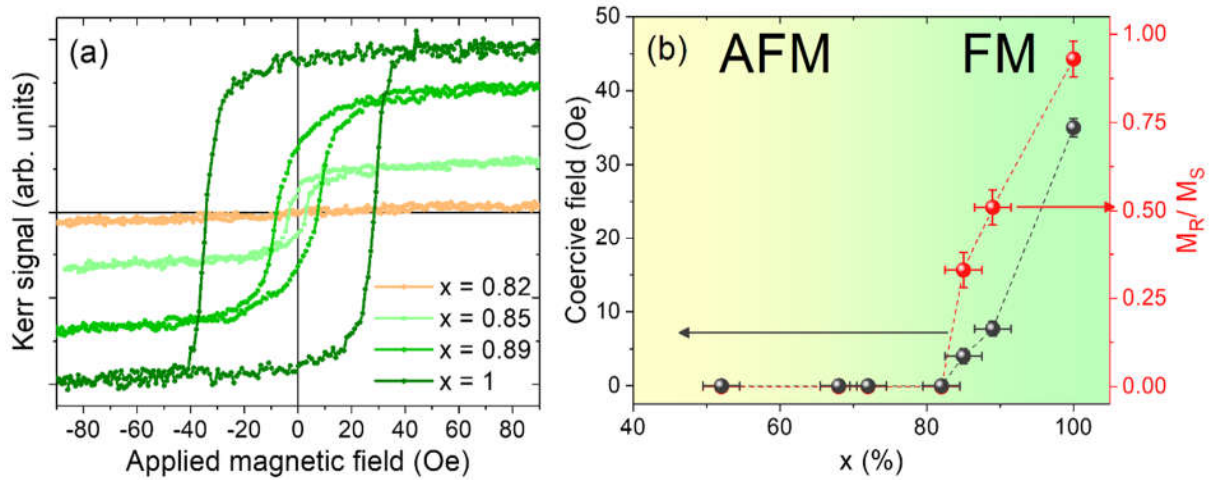

**Figure 2.** (a) MOKE hysteresis loop for x ratios going from 0.82 to 1. (b) Evolution of magnetic coercive field (black) and remanent magnetization (red) as a function of Fe x ratio in Fe$_x$Mn$_{1-x}$ thin film, going from AFM to FM regimes.

To define the x ratio at which the AFM/FM transition takes place, we carried out Kerr effect measurements on different samples. Figure 2 shows the evolution of magnetic hysteresis loops as a function of Fe$_x$Mn$_{1-x}$ x ratio for pristine substrates. For x ≤ 0.82, no signs of hysteresis is present (Figure 2a), indicating an overall AFM regime of the FeMn layer. For larger x value hysteresis appeared (FM regime), with increasing magnetic coercivity and remanence in correspondence of larger Fe concentrations. The evolution of these two parameters is plotted in Figure 2b. Such value of AFM/FM transition (around 0.85) is in good agreement with what reported on similar systems.[41,44]

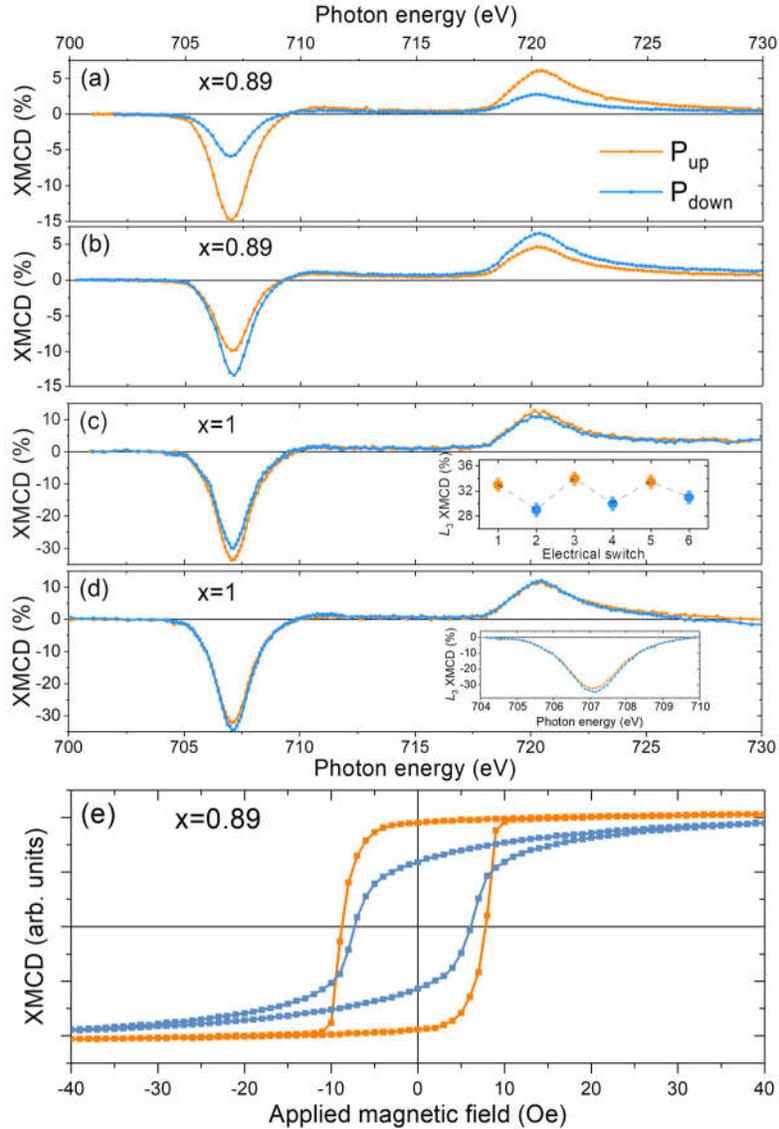

**Figure 3.** XMCD at Fe $L_{2,3}$ edges for x = 0.89 (a-b) and x = 1 (c-d) after positive $P_{up}$ (orange) and negative $P_{down}$ (blue) electrical switches. In the inset of (c), Fe $L_3$ edge values after subsequent electrical switches; (e) XMCD hysteresis loop at Fe edge for x=0.89.

In order to investigate the effects of converse magnetoelectric coupling, we took x-ray absorption spectroscopy (XAS) and x-ray magnetic circular dichroism (XMCD) measurements of $Fe_xMn_{1-x}$ samples at different relative concentrations after both electrical switches. Fe and Mn $L_{2,3}$ edges were probed at the APE-HE beamline of the Elettra synchrotron radiation facility in Trieste.[45] All measurements were performed at room temperature, in total electron yield (TEY) mode, by recording the drain current through the gold wire of the top contact (Figure 1a), measuring at remanence after in-plane magnetic saturation of the sample. From the chemical point of view, XAS spectra probed that all samples had purely metallic Fe and Mn edges, with relative signal intensities in agreement with the stoichiometry of the samples. The XAS spectra maintained the pure metallic shape in all electrical states of the PMN-PT substrate, ruling out any oxidation during growth or after air exposure and, more interestingly, any ion migration from/to the substrate during the ferroelectric cycles. In-operando XAS and XMCD spectra were taken at zero voltage, after setting the polarization to the state $P_{up}$ or $P_{down}$.

Figure 3 shows the evolution of XMCD spectra as a function of electric polarization at Fe $L_{2,3}$ edges for x = 0.89 and x = 1 in different points of the samples. Firstly, in the pristine Fe sample, we measured $L_3$ XMCD intensity of 32 ± 1%, in agreement with what reported in literature for pure Fe thin films,[46,47] proving a full magnetization of the pure Fe layer in our measurement conditions. The presence of small percentages of Mn gradually reduces the intensity of the dichroic signal at Fe edge until samples with x ≤ 82%, which showed no dichroic signal, in agreement with what observed by Kerr effect (Figure 1b). Quite interestingly, Fe and Mn dichroic signals showed an antiparallel alignment in the FM regime, with Mn XMCD having a much smaller signal compared to the Fe one. Thus for high Fe concentrations we can consider the Mn atom acting as an impurity in the ferromagnetic host, reducing Fe magnetic moment. The reduction of the Fe dichroic signal for increasing Mn concentrations is in agreement with what reported for FeMn thin films on MgO,[41] where on the other hand Mn is reported to have weak parallel alignment.[44,48] The $Fe_xMn_{1-x}$ phase diagram is complex and very sensitive to chemical and structural environment,[49,50] thus the growth on PMN-PT may be at the origin of such AFM coupling between Fe and Mn in the FM regime. However, the precise determination of the complex magnetic structure of the FeMn around the critical point goes beyond the scope of this paper; from here on, we will therefore focus on the evolution of the Fe FM response as a function of the sample electric switch.

Figure 3a-b show the Fe dichroic spectra in two different zones of the $Fe_{89}Mn_{11}$ sample, after in-situ electrical switching. We can clearly see how, according to the chosen point of measure, passing from $P_{down}$ to $P_{up}$ states can lead either to a huge increase (Figure 3a) or a clear reduction (Figure 3c) of the dichroic signal. In one of the measurement points (Figure 3a) the electrical switch led to a giant variation of the dichroic signal of a factor 2.5. To such variation of XMCD at remanence, it corresponds a modification of the whole hysteresis loop, as shown in Figure 3e. The changes of both coercive field and remanent magnetization are signs of a possible modification of the anisotropic axis induced by the change of polarization state.

Such findings are extremely surprising for two main reasons: (a) the magnitude of the observed effects is huge, arriving to modify the XMCD of 250% in the best case; (b) the variations appear to be sensitive to the zone probed during the measurement, whose area is ~150 μm$^2$, rather than to the ferroelectric state of the substrate. The latter observation implies that the models generally used for explaining converse magnetoelectric coupling do not apply in this case. Firstly, charge screening length in the case of metallic layers is confined at the first atomic layer at the interface,[51] while our sample thickness of 4 nm overpasses purely interfacial effects; moreover, charging effects are expected to be uniform in case of epitaxial interface. Regarding ion migration, its presence would entail variations on the chemical composition of the metallic layer,[52,53] which are not detected by XAS spectra after electrical switches at either Fe or Mn edges. Finally, since the electric poling is out of plane, purely strain-driven effects would lead to a symmetric variation after complete polarization,[16,29,31] which is clearly not the case for our system. This last point highlights a very intriguing aspect of the observed effect: no known magnetoelectric coupling mechanism is able to explain a difference in the sign of the coupling through the surface of the same sample.

In order to verify if such a behavior is characteristic of the proximity of FeMn layer to the AFM/FM transition, we repeated local in-operando XMCD investigations on the pure iron sample. Whereas in the pristine case the Fe film showed full magnetization, the effect of electrically polarizing the sample led to local variations of the dichroic signal intensity, similarly to what observed in the case of x = 0.89, but with reduced intensity. Indeed, we measured variations of $L_3$ XMCD intensity which span from an increase of a factor $1.12 \pm 0.03$ passing from $P_{down}$ to $P_{up}$ states (Figure 3c), or a very small reduction (Figure 3d). Moreover, we observed a complete reversibility of the magnetic response after different polarizations, as shown in the inset of Figure 3c. We attribute to the robustness of the ferromagnetism of a pure Fe layer (compared to the $Fe_{89}Mn_{11}$ layer close to the AFM/FM transition) the reason of the reduction of the observed effect; however, the persistence of local variations of magnetoelectric coupling confirms that a new coupling mechanism is acting in our system.

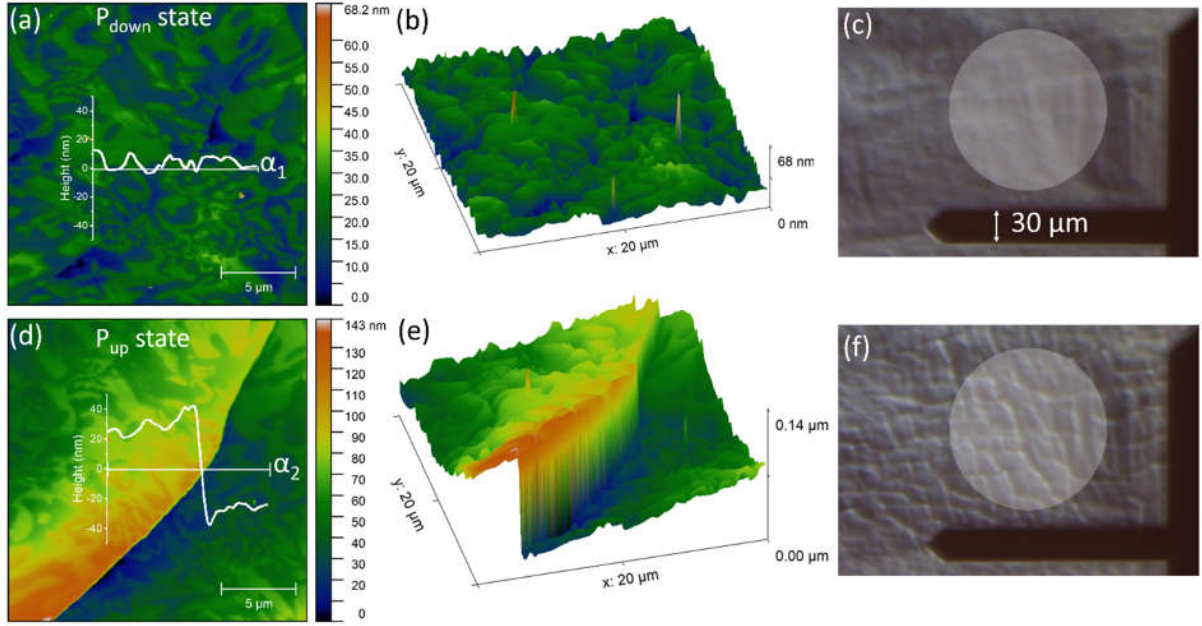

**Figure 4.** Atomic force microscopy topographic height images of a 20 x 20 μm² area of MgO/Fe$_{0.85}$Mn$_{0.15}$/PMN-PT sample after negative (a,b) and positive (d,e) electrical switching. A surface crack appears across the area; α$_1$ and α$_2$ show the topographic profiles in (a) and (d) scan over the same line of the sample. (c) and (f) are pictures from optical microscope of sample surface after negative (c) and positive (f) electrical switches. A tip width is used as size reference. Faded circles sign XAS/XMCD probed area per measurement.

To fully characterize our system we therefore decided to measure the surface topography as a function of the substrate polarization state; atomic force microscopy measurements were taken on the samples as a function of the electrical switch. Figure 4 shows the topographic evolution in the two different ferroelectric state P$_{down}$ and P$_{up}$ of PMN-PT substrate. In order to measure the topographic variations of the sample, the sample was measured at zero voltage after a negative (Figure 4a-c) and positive (Figure 4d-e) electrical switch, on the sample holder setup shown in Figure 1a.

The PMN-PT(001) substrates present a rich surface morphology, which is transferred to the thin layers deposited onto them. The surface is covered by island-like areas of different dimensions, ranging from tens to fractions of μm width. The height of the islands (in green in Figures 4a and

d) with respect to the background (in blue) is of few nanometers. This can be observed from the $\alpha_1$ profile measured in Figure 4a, with height variations of the order of tens of nm in correspondence to the passages from green to blue areas. The same topography was also measured on pristine PMN-PT(001) substrates, indicating that such morphology is not caused by a non-homogeneous deposition of $Fe_xMn_{1-x}$ layer, but it is directly due to the PMN-PT(001) substrate.

In addition to such morphology, the sample presents a second mesoscopic morphological feature, whose characteristics are sensitive to the electrical history of the sample. Indeed, Figure 4a presents a quite faint line crossing diagonally the scan from top right to bottom left, whose depth is of the same order of magnitude of the island morphology, as it can be seen in the $\alpha_1$ profile line crossing it. On the contrary, after positive electrical switch the same line transforms into a surface crack (see Figure 4d), whose height overpasses of a factor 10 the one present in the opposite electrical state. Profile $\alpha_2$ in Figure 4d, which crosses the same line as $\alpha_1$ in Figure 4a, shows the height of the surface crack that appeared after positive electrical switch. The amplitude of the morphological variation can be appreciated from the 3D view of the topological scans of Figure 4b and 4e.[54] We measured similar formations of surface cracks in several parts of the surface, with variations of heights as a function of electrical switch comparable with what shown in Figure 4. These variations were even visible by the optical microscope of the atomic force microscopy instrument. Figure 4c and f show the optical microscopy view of the surface after negative and positive electrical switches respectively. A complete modification of the surface morphology takes place all over the sample. Starting from the initial condition before biasing, the sample passes through a negative electrical switch, which modifies the sample morphology into a "smoother" one in correspondence to the negative current peak (Figure 4b). When the positive voltage is applied, the morphology abruptly switches to a corrugated one (Figure 4e) in correspondence to the positive current peak. We repeated the electrical switches several times, observing that the sample morphology reversibly passes from one to the other one. Surface cracks as those shown in Figure 4d-f vanish into smooth profiles all over the surface, with no signs of fatigue after several electrical transitions.

Such impressive electrically-driven morphological features are extremely intriguing from both the ferroelectric and the magnetoelectric points of view. Regarding the former one, it is known that ferroelectric/elastic materials tend to be fragile;[55] during ferroelectric transitions the crystal is subject to huge strain across its thickness, which may lead to substrate breakdown.[25] Moreover, cracks formed during cycling tend to expand after several repetitions.[56,57] In the case of PMN-PT, electrical switching corresponds to a 109° rotation of the ferroelectric domains,[30] passing through complex domain structures during polarization reversal.[58] Very recently, similar morphological modifications have been reported by Liu *et al.*[43] in the case of MnPt(35 nm)/PMN-PT(001) heterostructures. In their work, they show how the cracks induce a variation of the resistance of the metallic layer of several orders of magnitude as a function of the electrical switch, with endurance and perfect reversibility for up to $10^7$ cycles. The mechanism ruling the electrically-induced morphological transition in the two cases is similar: the $Fe_xMn_{1-x}$ layer, as the MnPt of Liu *et al.*,[43] acts as top electrode and follows the morphological variations of the substrate, which transfers onto the magnetic layer its formation and annihilation of cracks. One remarkable aspect present in both studies is that the metallic layer completely recovers the electronic properties (electrical in ref.44, magnetic in our case) with perfect reproducibility and remarkable endurance after the morphological transitions.

This aspect leads to the consequences of the mesoscopic morphological transformations onto the converse magnetoelectric coupling of our heterostructure. To be responsible of the observed changes in the XMCD signal, the surface cracks must have a high spatial density. In figure 4c and f, the grey circles show the dimension of the probed area during XMCD measurements. As it can be clearly seen in the figures, such area is large enough to cover a large number of surface cracks, i.e. the morphological state of two XMCD measurements is dramatically different on a large portion of the probed area.

As known, local defects and dislocations play an important role in magnetism, acting either as nucleation points for domain wall or at the opposite as pinning points during magnetization reversal; at remanence, they can modify both the coercive field and the total magnetic moment by locally creating non-uniform magnetization or spin glass states.[47,59] In our case, the ferroelectric domain reversal leads to complex modifications across the thickness of the sample due to the structural changes taking place during the polarization reversal. As shown in Figure 4, we have observed that after positive polarization such processes lead to the formation of cracks, which create areas of tens of µm$^2$ separated one from the other by vertical walls of almost a hundred nm height. Despite the dimension of such areas is too large to allow single domain rotation during magnetization reversal instead of domain wall nucleation and propagation,[60] cracks separate zones which previously were magnetically connected. If in this condition the electrical polarization is reversed with a negative electrical field, areas previously vertically separated return at similar height because of the transition of surface cracks into smooth lines, recreating the magnetic connections across it. As shown in Figure 3, the magnetic responses of the thin films are sensitive to the morphological variations, recreating magnetic interactions along zones that were previously magnetically uncorrelated. XMCD spectra were taken at remanence after magnetic in-plane saturation, which implies that the modulations of dichroic signal intensity after electrically-driven morphological transitions are due either to a variation of the magnetic anisotropy, i.e. of the remanent magnetization, or to a change of the total magnetic moment, i.e of the net magnetization. The reproducibility of the magnetic responses after morphological modifications shows that such changes of the magnetic interactions can be reversibly broken and reformed, despite mechanical fractures take place on the surface of the sample and across the FeMn thin film. Since the density of cracks formed after electrical switch is randomly distributed over the sample surface, its effects on the magnetic response varies randomly according to the probed zone. For this reason, the amplitude of the converse magnetoelectric effect on a probed area is affected by the local morphology, spreading from huge coupling effects to none.

These findings add a new player in the interfacial coupling of artificial heterostructures, whose amplitude, as it is the case for our study, can overpass the one attributed to charge or strain effects. Also for our case of study, we cannot exclude the presence and contribution of other coupling mechanisms in our system. As previously stated, on one side charges are accumulated at the interface with $Fe_xMn_{1-x}$ after the current peak measured in I(E) curves, and on the other one polarization reversal of PMN-PT creates large strains at ferroelectric boundaries because of the 109° angle of polarization switching between domains. Ongoing studies of the relative weight of these aspects are under investigation. Nonetheless, such effects do not explain the huge variations of magnetoelectric response depending on the measurement zone on the sample, which arrive to invert completely the observed effect, thus demonstrating the dominant role of the morphology in this system. This new mechanism of electrically-induced variation of magnetic response offers a wide range of possibilities through which the morphological driven magnetoelectric coupling could be maximized, such as using magnetostrictive materials (FeGa or Ni) as a metallic layer or patterning the ferroelectric substrate to gain control on the induced modifications.

In summary, we have reported how $Fe_xMn_{1-x}$/PMN-PT samples experience electrically induced reversible morphological changes, with the formation and annihilation of surface cracks of tens of nm depth. Such morphological modifications of the substrate are transferred from the PMN-PT substrate to the thin $Fe_xMn_{1-x}$ layer, leading to huge and reversible modifications of its ferromagnetic response. These findings add a new degree of control of magnetism driven by an electrical field and make surface morphology an additional important parameter that opens new promising possibilities in the design of new multiferroic heterostructures with enhanced performances.

**METHODS**

**Sample deposition**

$Fe_xMn_{1-x}$ thin films of 4 nm thickness were deposited by MBE co-deposition from Fe and Mn crucibles on (001)-oriented PMN-PT single crystal substrates (5 x 5 x 0.5 $mm^3$) at room temperature, with a base pressure of 1 x $10^{-9}$ mbar. x ratio was modulated by varying the deposition rates of the two evaporators, which were calibrated by quartz microbalance. Deposition rate was fixed at 1.9 Å/min for Fe evaporator and ranged from 0.19 to 1.9 Å/min for Mn evaporator. Samples were then capped by 3 nm thick MgO layer with 0.57 Å/min deposition rate. All substrates were in pristine state during deposition. PMN-PT single crystal substrates were purchased from SurfaceNet GmbH.

**MOKE characterization**

Longitudinal Kerr effect measurements were done at ambient pressure, room temperature condition using a *s*-polarized blue laser (435 nm) and intensity modulation at 815 Hz. All measurements have been taken ex situ after MBE deposition. Laser spot size is about 500 μm$^2$. Coercive field is calculated as hysteresis loop half width at zero magnetization.

**Electrical characterization**

Electrical contacts onto FeMn film were made via contacting a gold wire on top of the surface and using a conductive silver paint contact on the bottom part. The two electrically isolated contacts of the sample holder were connected to a Keithley 6485 picoammeter /voltage source at room temperature. Measurements were taken both in UHV conditions and in air, with no changes in the electrical response. I(E) curves shown in the article were taken with a NPLC of 1.

**Morphological characterization**

Atomic force microscopy measurements were acquired by A100 microscope of A.P.E. Research. Cantilevers with stiffness of 40 Nm and lenght of 125 micrometers were used. The sample was measured at ambient pressure at zero voltage, after cycling the electric field to positive or negative saturation using the sample holder connections in the measurement position. Tip was lifted during the application of the bias to avoid surface damaging during the morphological transitions.

**XAS/XMCD characterization**

All measurements were performed at room temperature, in total electron yield (TEY) mode, normalizing the intensity of the sample current to the incident photon flux current at each energy value. Absorption spectra were taken in circular polarization, with the sample surface at 45° with respect to the incident beam. In this condition, the footprint of the x ray on the sample surface covers an area of ~150 μm$^2$. XMCD spectra were recorded at remanence, alternating magnetic field pulses of ±300 Oe in the surface plane at each energy point of the spectra, to saturate sample magnetization. The spectra are presented normalizing to 1 the sum of the intensity of the $L_3$ lines. The difference between the two resulting curves gives the dichroic signal; XMCD signal expressed in % takes into account the angle of 45° between the photon angular momentum and the sample magnetization, as well as the 75% circular polarization degree of light.

**Corresponding Authors**

Giovanni Vinai

E-mail: vinai@iom.cnr.it

ORCID: https://orcid.org/0000-0003-4882-663X

Piero Torelli

E-mail: piero.torelli@elettra.eu

**Author Contributions**

The manuscript was written through contributions of all authors. All authors have given approval to the final version of the manuscript.

**Acknowledgments and fundings**

This work has been performed in the framework of the nanoscience foundry and fine analysis (NFFA-MIUR Italy Progetti Internazionali) project. We acknowledge Alessio Guerra for the video editing of the supplementary information.